\documentclass[conference]{IEEEtran}
\usepackage{blindtext, graphicx}
\ifCLASSINFOpdf
\else
\fi
\usepackage{caption}

\usepackage{pgfplots}

\pgfplotsset{compat=1.5}

\usepackage{listings}

\usepackage{csquotes}
\usepackage[english]{babel}
\usepackage{biblatex}
\begin{document}
%
\title{CGui Toolchain for Highly Customized GUI Development for Multiple Platforms}

\author{\IEEEauthorblockN{Samiyuru Menik}
\IEEEauthorblockA{Department of Computing \\
Informatics Institute of Technology \\
Affiliated with University of Westminster UK \\
Email: samiyuru@gmail.com}
\and
\IEEEauthorblockN{Sriyal Jayasinghe}
\IEEEauthorblockA{Department of Computing \\
Informatics Institute of Technology \\
Affiliated with University of Westminster UK \\
Email: sriyal@iit.ac.lk}}


%


\maketitle

\begin{abstract}
Highly customized graphical user interfaces play a major role in today's software applications. There exist many technologies that support developing them. But these technologies demand writing a significant amount of unintuitive code to implement customized graphical user interfaces. Also, the support of these technologies become very limited when developing user interfaces for multiple platforms. In this paper we present a new technology to develop highly customized graphical interfaces in a more effective and efficient method. In the first part of this paper we explore the existing graphical user interface development technologies and points out their strengths and limitations. Then we introduce CGui toolchain addressing the identified limitations. Finally, we evaluate the effectiveness of CGui toolchain in comparison to two existing representative technologies that supports highly customized graphical user interface development. 
\end{abstract}


\IEEEpeerreviewmaketitle

\section{Introduction}

Graphical User Interface (GUI) is an integral part of a software product. The GUI of a software system is the main access point where users interact with the whole system. Therefore, GUI determines a large part of the success or the failure of a system \cite{Zhao2012}. Further, the current trend of GUI is not to focus only on the minimal that is required to complete a task. But they have to be attractive, feature rich and usable across a variety of different computing devices to win the competitive market. Developing GUI to meet these standards is expensive and challenging \cite{Cerny2012}.
Developing a customized GUI for a given platform requires an in-depth knowledge and practice in several technologies. It requires to know the targeted platform, its GUI development framework, the supported programming language and the related tools \cite{Kulloli2013}. Thus, if an application is targeted on multiple platforms, above knowledge requirements should be met for all the platforms \cite{Scoditti2009asd}. This consumes a large amount of time and resources that could otherwise have been invested in developing application features.

\begin{figure}[h]
	\centering
	\includegraphics[width=0.7\linewidth]{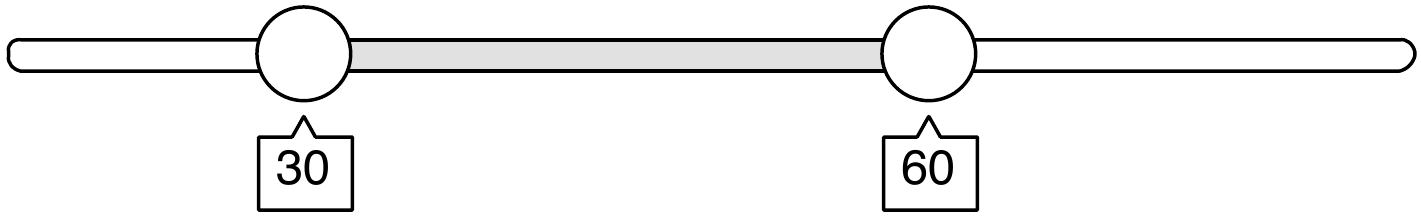}
	\caption{Custom Range Slider}
	\label{range-slider}
\end{figure}

With the current GUI development technologies, developers are confined to a set of abstract concepts such as layout managers that are embedded into the technologies. Even though these concepts are intended to make solving common problems easier, they fail to support developing unorthodox solutions. As an example consider the custom range slider in Fig. \ref{range-slider}. In this slider the space between the movable nodes should be marked. The two tool tips should move with their corresponding nodes. The text of the tool tip should show the value of the related node which is a function of the current position of the node. A behavior and a set of layout relationships that is customized to this extent is impossible to directly map to a set of abstract concepts provided out of the box by a GUI development technology. Developing GUI like these require the developers to write a lot of technology specific code. In this case, developers have to pay more attention to the implementation details rather than focusing on the quality of the output \cite{luostarinen2010user} leading to an increased development cost and a higher possibility of bugs. Furthermore, as the number of target platforms and devices grows, these complications also grow proportionally \cite{debbarma2012static}.

The rest of this paper is organized as follows: In Section II we explain the set of qualities that are expected from a GUI development technology. In Section III, we briefly discuss the qualities of the approaches of GUI development technologies. In Section IV, we discuss the design of CGui toolchain. In Section V we evaluate the effectiveness of CGui toolchain and Section VI concludes.

\section{Success Factors of a GUI Development Technology}

Researchers have pointed out a number of qualities that should be there in a successful GUI development technology. By considering them, following key success factors \cite{heitkotter2013evaluating, dalmasso2013survey} have been derived considering the focus of this project that emphasizes the productivity and convenience of customized GUI development.

\noindent
\textbf{Control Over The GUI} -- This represents the amount of details in the GUI that the technology allows to control. The technology should support the developers to implement more customized GUI taking over the control of development concerns \cite{6484050} in a higher granularity.

\noindent
\textbf{Generality} -- This represents applicability of the GUI that are developed with a specific technology. For instance, a GUI development technology should be capable of producing GUI for multiple devices and platforms with different aesthetic and usability properties \cite{akiki2015adaptive}.

\noindent
\textbf{Impact on Development and Maintenance} -- The GUI development technology should provide features to easily develop the GUI. Also the GUI developed using the technology should be easily maintainable.

\noindent
\textbf{Learning Curve} -- This represents how hard its is to learn the technology in order to develop a GUI utilizing full power of the technology.
 
\noindent
\textbf{Technology Independence} -- The GUI development technology should be able to produce GUI that are compatible with a wide range of existing technologies. And it should be adaptable to the technologies that can emerge in the future. This also implies the possibility of deploying the GUI in multiple platforms that supports different technologies.

\section{GUI Development Approaches}

Different GUI development technologies can be categorized by the approaches they follow to support GUI development.

\subsection{Object Oriented GUI Frameworks}

These frameworks provide a standard set of reusable GUI widgets on top of an object oriented programming language to compose a GUI \cite{bederson2004toolkit}. Then event handlers \cite{bishop2004developing} are used to control the behavior of the application.

\noindent
\textbf{Control over the GUI} -- Higher control over the GUI due to the exposure of low level details through object oriented API.

\noindent
\textbf{Generality} -- Depending on the implementation of the framework, generality can be achieved.

\noindent
\textbf{Impact on development, maintenance} -- Hard to develop and maintain in comparison to the alternative approaches mainly due to the exposure of low level implementation details to the developers \cite{appert2008swingstates}.

\noindent
\textbf{Learning curve} -- Difficult to learn due to the required knowledge in internals of the framework as well as the advanced programming techniques to develop customized GUI.

\noindent
\textbf{Technology independence} -- Not independent. Depends on the underlining programming language.

\subsection{Widget Based GUI Builders}

GUI builders are visual development tools that are built on top of object oriented GUI frameworks. They allow to design and develop GUI by composing a set of reusable widgets in a WYSIWYG manner using visual direct manipulation techniques \cite{zeidler2013auckland}.

\noindent
\textbf{Control over the GUI} -- Widgets can be customized only by the provided configuration options through the GUI builder. Also the implementation is limited by the specification capability in the visual composition techniques \cite{nebeling2014interactive}.

\noindent
\textbf{Generality} -- Depends on the underlining GUI development framework.

\noindent
\textbf{Impact on development, maintenance} -- The development is relatively easy since the GUI builders automatically generate code for the GUI. Also, the WYSIWYG nature simplifies the GUI development. But the maintenance can be hard because of the generated unreadable code.

\noindent
\textbf{Learning curve} -- Relatively easy to learn because of the reduced amount of code that needs to be written for the GUI.

\noindent
\textbf{Technology independence} -- Depends on the underlining GUI development framework and its programming language.

\subsection{User Interface Description Languages}

User Interface Description languages (UIDL) are a form of a declarative GUI specification languages that are mostly XML dialects. In order to render a GUI specification of these languages generally a rendering engine required \cite{guerrero2009theoretical}.

\noindent
\textbf{Control over the GUI} -- Depending on the semantics of the UIDL it can provide a good control over the structure and the layout of the GUI. But the UIDL cannot define the behavior of a GUI. Because of that the control over the GUI is limited.

\noindent
\textbf{Generality} -- UIDL can separate the GUI definition from its rendering \cite{vanderdonckt2004usixml}. This allows to render the GUI dynamically to suit the target platform. Therefore generality can be achieved.

\noindent
\textbf{Impact on development, maintenance} -- UIDL simplifies the GUI development and maintenance by reducing the requirement to use imperative programming languages to define the GUI. Also this reduces the possibility of programming errors \cite{souchon2003review}.

\noindent
\textbf{Learning curve} -- Most UIDL are XML based simple human readable languages. Therefore these languages do not require a deep technical expertise. But generally the vocabularies of these languages are complex and the their learning curve is steep.

\noindent
\textbf{Technology independence} -- UIDL can be rendered using any technology that implements the language specification. Therefore, UIDL are technology independent.

\subsection{Model Based and Automatic Techniques}

Goal of this approach is to generate plastic user interfaces for multiple targets using several abstract layers of higher level models without dealing with low level GUI programming details \cite{meixner2011past}.

\noindent
\textbf{Control over the GUI} --  Existing languages that are used for modeling in this approach are not flexible enough to provide a good control over the GUI \cite{akiki2015adaptive}.

\noindent
\textbf{Generality} -- Multi-targeting and plasticity is achievable with a reduced amount of duplicate work \cite{Coutaz:2010:UIP:1822018.1822019}.

\noindent
\textbf{Impact on development, maintenance} -- In model based UI development low level implementation details are hidden from the developers. Only the high level models are exposed to them. Therefore development and maintenance of GUI developed in this approach is relatively easy.

\noindent
\textbf{Learning curve} -- Model based techniques require number of specialized models and related tools in order to derive the final GUI. Because of that model based techniques are hard to learn \cite{meixner2011past}.

\noindent
\textbf{Technology independence} -- Since the platform specific GUI are automatically generated from technology independent models, the approach is technology independent.

When considering the above approaches, all techniques except object oriented GUI frameworks provided a level of abstraction to hide the complexities of the low level implementation details of GUI development. But those approaches had to compromise a degree of control over the GUI to provide that abstraction. On the other hand, the GUI preview feature of GUI builders, the declarative nature of UIDL and the code generation techniques employed in model based techniques can be identified as positive aspects of the existing approaches.

\section{Design of CGui}

With CGui our goal is to come up with a solution that harness the strengths of existing GUI development technologies and address the recognized limitations. Therefore, we define a set of qualities for CGui to optimize the previously mentioned success factors of a GUI development technology.

\noindent
\textbf{Control over the GUI} -- GUI is essentially a set of relationships between the elements of GUI and the underlining program. Therefore, with the help of previous work \cite{jamil2014constraint, zeidler2012comparing, Lutteroth:2008:DSH:1394991.1394993}, we identify mathematical constraints as a powerful method for defining GUI. By using constraints to define the GUI, CGui can provide a greater control over the GUI in a declarative fashion and make the definition of the GUI simpler.

\noindent
\textbf{Generality} -- CGui has an extensible set of runtime libraries that are responsible for rendering the GUI in the most optimized way for a given target platform.

\noindent
\textbf{Impact on development, maintenance} -- CGui language supports modularising the GUI into smaller reusable components to make development and maintenance convenient. Also, CGui has a sophisticated integrated development environment (IDE) to assist in developing the GUI.

\noindent
\textbf{Learning curve} -- CGui contains a small number of intuitive constructs that are easier to learn.

\noindent
\textbf{Technology independence} -- CGui has an extensible set of cross compilers to generate GUI modules for each platform without compromising performance. Because of these cross compilers, the GUI developed in CGui can be coupled with many technologies without modifications.

\subsection{CGui Development Process}
\begin{figure}[h]
	\centering
	\includegraphics[width=\linewidth]{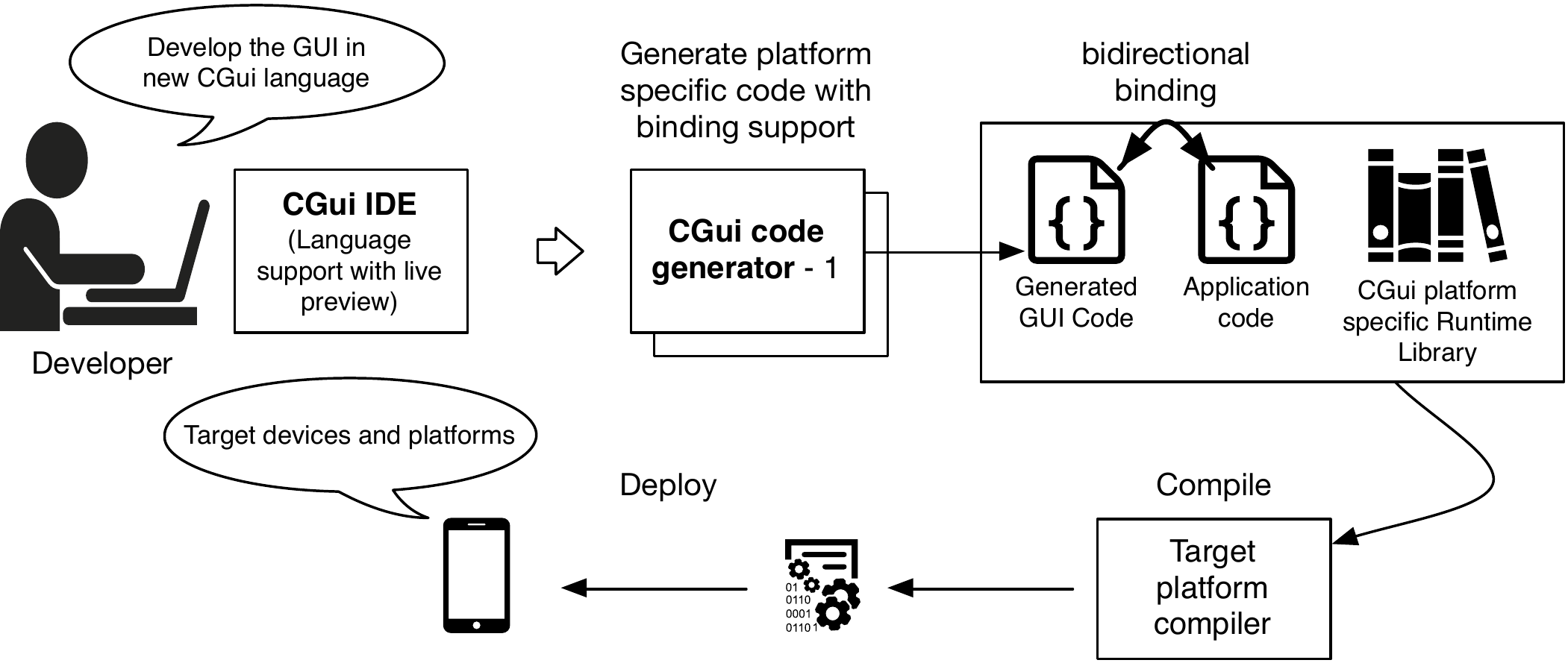}
	\caption{CGui development process}
	\label{cgui-workflow}
\end{figure}

Fig. \ref{cgui-workflow} shows the steps between specifying a GUI with CGui and deploying it in target devices after the data binding with the application code.

\subsection{CGui Language}
	
Listing \ref{cgui-fragment} shows a part of a CGui code block.

\lstset{
	caption=CGui Code,
	label=cgui-fragment
}
\begin{lstlisting}
@gui
  // GUI structure
  dialog{ ok cancel }
@constraints
  // Property relationships
  dialog.W << base.W   
  dialog.H << base.H
\end{lstlisting}

\subsubsection{\textbf{Structure of a CGui Code file}}
A CGui code file consists of three major parts; GUI structure, property constraints and variable exports. GUI structure specifies the parent child relationships of the GUI elements. Property constraints define the relationships between the properties of the GUI elements and the variables. Variable exports declare what variables should be exposed to the outside world when the GUI module is generated.

\subsubsection{\textbf{Constraints}}
\begin{figure}[h]
	\centering
	\includegraphics[width=0.7\linewidth]{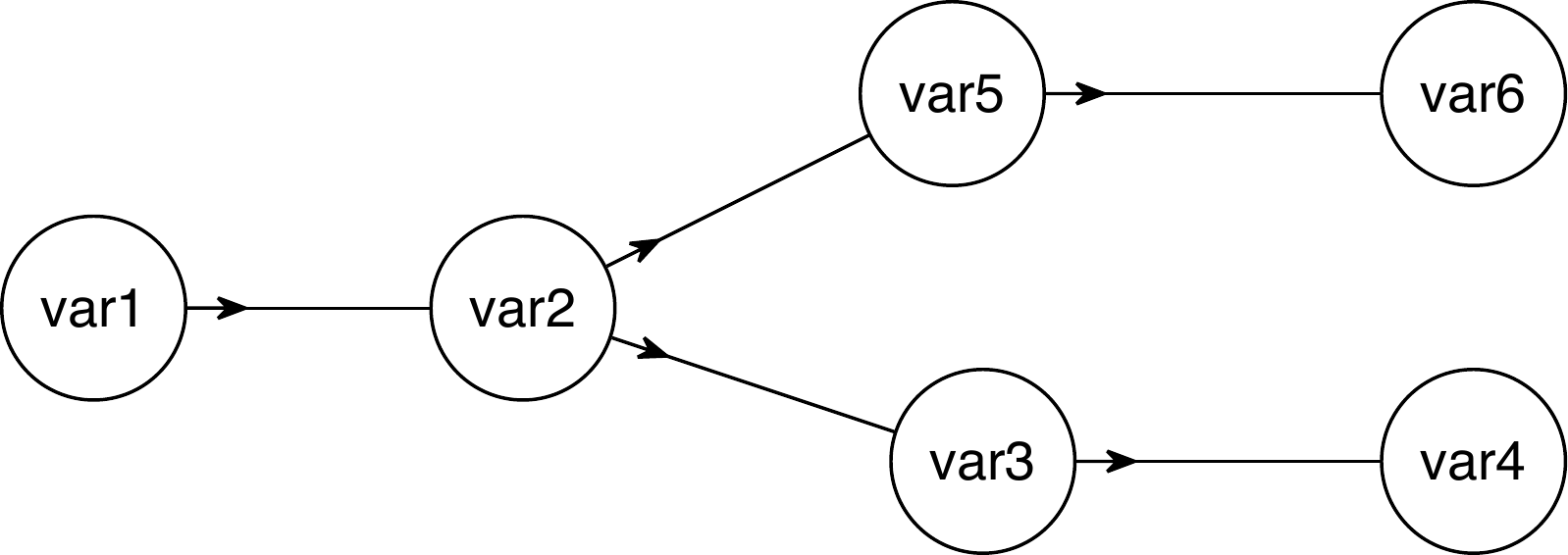}
	\caption{Directed Variable Value Propagation Graph}
	\label{dircted-var-graph}
\end{figure}

CGui language supports non linear one-way equality constraints and inequality constraints between properties of the GUI and other variables. In order to maintain the one way relationships, a constraint solver that utilizes a directed variable value propagation graph is used. See Fig. \ref{dircted-var-graph}. To avoid possible cycles when propagating values in this graph, a mark sweep mechanism is used. In order to maintain the inequality relationships, once the constraints are solved after a change of a value, decision of applying the new values to the system or discarding them is made only after checking the inequality constraints that has become outdated due to the value change and the subsequent constraint solving phase.

\subsubsection{\textbf{CGui Composable Function Modules}}
CGui has an extensible set of reactive functions to allow specifying relationships that are hard or impossible to specify only with arithmetic operators. These functions can be composed to create further complex relationships in the GUI. Math.min(x, y) and Math.max(x, y) are examples for these functions. Here, when the values of x and y variables change with other constraints, Math.min will update its return value to the minimum from x and y values and Math.max will update its return value to maximum from x and y values. These functional relationships blend with the general arithmetic relationships to create the complete graph of relationships in the GUI.

\subsubsection{\textbf{Data Binding}}
CGui language provide @export var1, var2 ... construct to expose variables to the outside world. When variables are declared in this way, code generation will create an observable property for each exported variable in the output module. These properties can be used to get, set or listen to the values of the underlining CGui variables from the target programming language.

\subsubsection{\textbf{Modularity}}
CGui is a modular language. With CGui GUI can be broken into smaller components and implement them in independently in separate files. One CGui module can be imported into another CGui module with var1:ModuleName syntax to later compose the complete GUI. Properties of an imported module can be accessed using their fully qualified name and use them as if they are normal in module properties.

\subsection{CGui IDE}

\begin{figure}[h]
	\centering
	\includegraphics[width=1\linewidth]{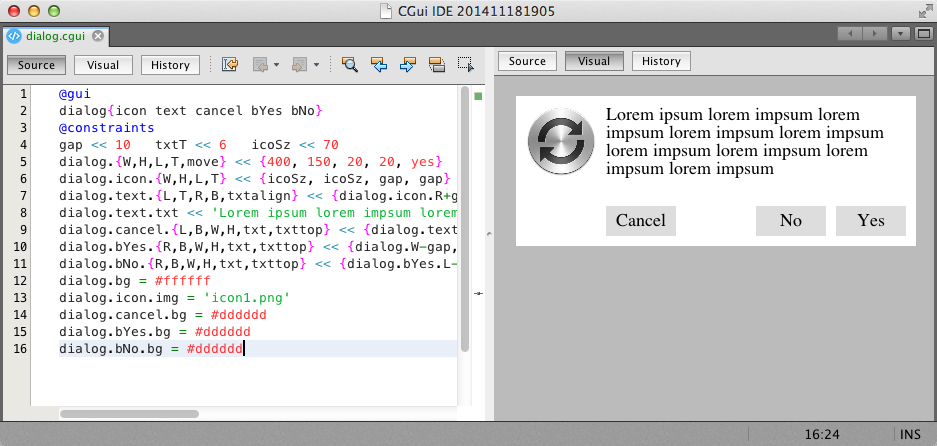}
	\caption{CGui IDE with live GUI preview support}
	\label{cgui-ide-dlgprvw}
\end{figure}

CGui IDE provides syntax highlighting, real-time conflicting constraint identification assistance and real-time interactive live preview for the GUI that are developed in CGui. See Fig. \ref{cgui-ide-dlgprvw}. The interactive live preview is an immediate feedback for the developers to see the effects of the changes they make to the GUI and validate the outcome. 

\subsubsection{\textbf{Code Generators}}
CGui IDE has a pluggable set of code generators that each code generator supports converting a CGui implementation to a specific programming language of a platform. This conversion produces GUI module that can be connected with the application code with bidirectional data binding.

\subsection{Runtime Libraries}
Each supported target platform of CGui has its own platform specific runtime library. This runtime library provides the required dependencies to run the generated CGui module on the target platform. These runtime libraries are also responsible of event dispatching and rendering the GUI on the target platform in the most optimized way. In addition to that, runtime libraries are capable of maintaining the consistency of the GUI across multiple platforms by mapping platform events to the matching CGui events and by ensuring consistent view rendering across platforms.

\section{Evaluation}

\subsection{Productivity and Learning Curve}

\begin{figure}[h]
	\centering
	\includegraphics[width=0.7\linewidth]{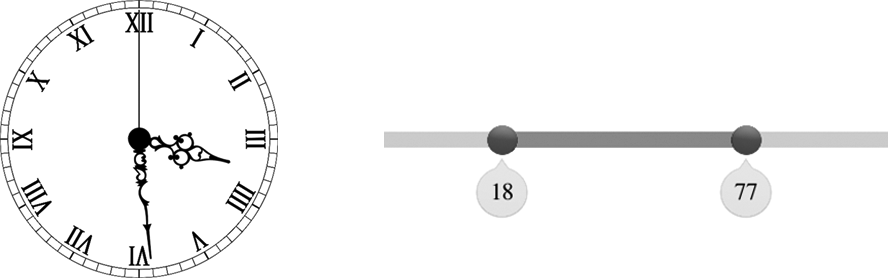}
	\caption{Prototypes of the working analogue clock (left) and the range-slider (right)}
	\label{clock-and-rangeslider}
\end{figure}

In order to test the learning curve and the productivity of CGui toolchain, an experiment was conducted with 6 Java developers and 7 HTML5 developers. They were provided with the two custom GUI prototypes that are shown in Fig. \ref{clock-and-rangeslider} along with all the resources required for the implementation. Then Java developers were told to develop the prototypes separately in Java and CGui. HTML 5 developers were told to develop the prototypes separately in HTML5 technology stack and CGui. Before starting the CGui implementation the developers were given a 20 minutes of training on CGui language and the IDE. Also they were given a short reference card for CGui language syntax. Finally, the time taken by each developers to implement the GUI prototypes were recorded.

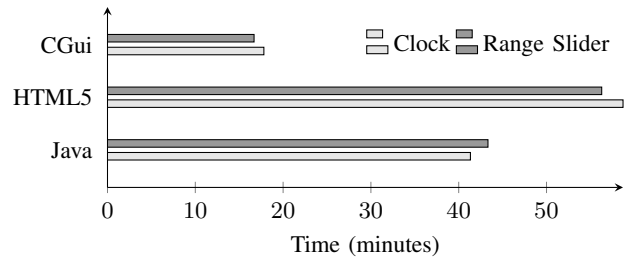
\begin{figure}[h]
\centering
\begin{tikzpicture}
    \tikzstyle{every node}=[font=\small]
	\begin{axis}[
			xbar,                                     
			y                 = 7mm,
			bar width         = 1mm,
			xmin              = 0,
			axis y line       = left,                 
			axis x line       = bottom,               
			xlabel            = Time (minutes),       
			x label style     = {anchor=north},       
			enlarge y limits  = {abs=5mm},            
			ytick             = data,                 
			ytick style       = {draw=none},          
			symbolic y coords = {Java,HTML5,CGui},    
			legend style      = {
                                  anchor=north east,  
                                  at={(1,0.91)},         
				                  draw=none,          
				                  legend columns=-1   
				                }
		]
		\addplot [fill=gray!20] coordinates {                        
			                   (41.3333333333333,Java) 
			                   (58.7142857142857,HTML5) 
			                   (17.8104395604396,CGui) 
			                 };
		\addplot [fill=gray!80] coordinates {                        
			                   (43.3333333333333,Java) 
			                   (56.2857142857143,HTML5) 
			                   (16.6923076923077,CGui) 
			                 };
		\legend{Clock, Range Slider}
	\end{axis}
\end{tikzpicture}
\caption{Average time spent by Java and HTML5 developers to develop the Clock and the Range slider}
\label{time-java-html5-dev}
\vspace{-3mm} 
\end{figure}

According to the test results, both Java and HTML developers were able to implement both prototypes with CGui in a significantly less time than with their familiar technology. See Fig. \ref{time-java-html5-dev}. Hence, it is evident that CGui is capable of significantly increasing the productivity of GUI developers when developing customized GUI. In addition to that, since all developers were able to implement the two GUI prototypes successfully in CGui with a training of only 30 minutes, it can be concluded that CGui is easy to learn.

\subsection{Lines of Code}

\begin{figure}[h]
	\centering
	\begin{tikzpicture}
	\begin{axis}[
			xbar,                                     
			y                 = 7mm,
			bar width         = 1mm,
			xmin              = 0,
			axis y line       = left,                 
			axis x line       = bottom,               
			xlabel            = Time (minutes),       
			x label style     = {anchor=north},       
			enlarge y limits  = {abs=5mm},            
			ytick             = data,                 
			ytick style       = {draw=none},          
			symbolic y coords = {Java,HTML5,CGui},    
			legend style      = {
                                  anchor=north east,  
                                  at={(1,0.91)},         
				                  draw=none,          
				                  legend columns=-1   
				                }			                
	]
	\addplot [fill=gray!20] coordinates {                        
		(120,Java) 
		(96,HTML5) 
		(28,CGui) 
	};
	\addplot [fill=gray!80] coordinates {                        
		(310,Java) 
		(157,HTML5) 
		(20,CGui) 
	};
	\legend{Clock, Range Slider}
	\end{axis}
	\end{tikzpicture}
	\caption{Average number of lines of code taken by Java and HTML5 developers to develop the Clock and the Range slider}
	\label{loc-java-dev}
	\vspace{-3mm} 
\end{figure}
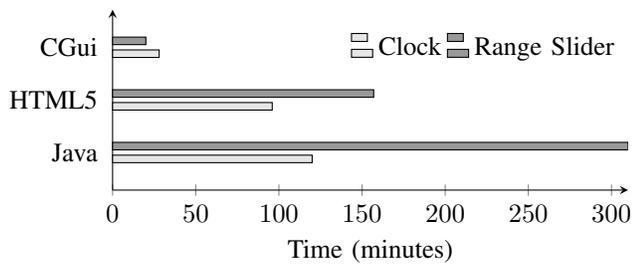

Using the implementations done by the developers in the above experiment, the average lines of code that were required to implement the GUI prototypes in fig. 3 using each technology were calculated. See Fig. \ref{loc-java-dev}. According to the test results, the developers were able to implement the two GUI prototypes in a significantly less number of code lines with CGui. Thereby, it can be assumed that one of the main reasons to the increased productivity in CGui was the less number of code lines required in CGui to implement the GUI.

\section{Conclusion}
We introduced CGui declarative constraint based language and the toolchain in order to develop highly customized GUIs for multiple platforms with a greater productivity and a significantly smaller learning curve. We directly embedded the constraints to the CGui language without introducing abstractions that would limit the generality and flexibility of the concept constraints. We believe this combined with the adequate support of tooling was the reason for the straightforwardness and the higher productivity resulted from CGui.

\subsection{Future Work}
CGui should have language constructs to specify GUI with collections of views. 
Also, in addition to the existing box view, CGui should have a method to add custom view types with custom properties. Further, CGui should extend its set reactive functions to provide more generic composable reactive mappings. Finally CGui should be further tested in comparison to other GUI development technologies with different GUI samples.

\printbibliography

\medskip

\noindent

\end{document}